\documentclass[letterpaper, 10 pt, conference]{ieeeconf}  

\usepackage{graphicx}
\usepackage{amssymb}

\usepackage{amsmath}
\usepackage{bm}
\usepackage{dsfont}

\usepackage{float}
\usepackage{mathrsfs}
\usepackage{subfigure}
\usepackage{times}
\usepackage{xspace}
\usepackage{epstopdf}

\newcommand{\gss}[3]{\mbox{\boldmath $#1$}_{#2}^{#3}}
\newcommand{\g}[1]{\mbox{\boldmath $#1$}}

\newcommand{\sbm}[1]{\mbox{\scriptsize $\g{#1}$}}
\newcommand{\sbms}[2]{\mbox{\scriptsize $\gss{#1}{#2}{}$}}

\newcommand{\pref}[1]{(\ref{#1})}

\newtheorem{prop}{\textbf{Proposition}}

\IEEEoverridecommandlockouts                              
\title{\LARGE \bf Estimator Selection: End-Performance Metric Aspects$^{*}$\thanks{Finalized version.}}

\author{Dimitrios Katselis, Cristian R. Rojas, Carolyn L. Beck 
\thanks{D. Katselis and C. L. Beck  are with the Coordinated Science Laboratory and the Department of Industrial and Enterprise Systems Engineering, University of Illinois at Urbana-Champaign,
Urbana, IL 61801-2925, Emails:
{\{katselis|beck3\}}{@illinois.edu}.}
\thanks{Cristian R. Rojas is with the ACCESS Linnaeus Center, Electrical Engineering, KTH Royal Institute of Technology, S-100 44 Stockholm, Sweden. Email:
                          {cristian.rojas}{@ee.kth.se}.}}

\begin{document}
\maketitle \thispagestyle{empty} \pagestyle{empty}

\begin{abstract}
Recently, a framework for application-oriented optimal experiment design has been introduced. In this context, the distance of the estimated system from the true one is measured in terms of a particular end-performance metric. This treatment leads to superior unknown system estimates to classical experiment designs based on
usual pointwise functional distances of the estimated system from the true one. The separation of the system estimator from the experiment design is done within this new framework by choosing and fixing the estimation method to either a maximum likelihood (ML) approach or a Bayesian estimator such as
the minimum mean square error (MMSE). Since the MMSE estimator delivers a system estimate with lower mean square error (MSE) than the ML estimator for finite-length experiments, it is usually considered the best choice in practice in signal processing and control applications. Within the application-oriented framework a related meaningful question is: Are there end-performance metrics for which the ML estimator outperforms the MMSE when the experiment is finite-length? In this paper, we affirmatively answer this question based on a simple linear Gaussian regression example.
\end{abstract}

\section{Introduction}

A basic subproblem in the context of system identification is that
of experiment design. Overviews of this topic over the last decade
can be found in
\cite{g05,h05,p08,h09}.
Contributions include convexification \cite{jh05}, robust
design \cite{krwh12,rwgf07}, least-costly design
\cite{bsghh06}, and
closed vs open loop experiments \cite{ag07}.

Recently, a new framework for performing experiment design has been introduced. This framework
is termed \emph{application-oriented experiment design} and it has been outlined in \cite{h09}. Specific investigations related
to communication systems were performed in \cite{krhb12,krbbbsjh13}. Denoting the end-performance metric by $J$ and assuming that
$J$ depends on the true and the estimated models, the performance is considered to be acceptable if $ J\leq 1/\gamma$ for some
parameter $\gamma$, which we call \emph{accuracy}. This motivates the introduction of a set of admissible models $\mathcal{E}_{adm}=\left\{G:J\leq 1/\gamma\right\}$, where $G$ denotes the model to be inferred. With these definitions, the least-costly experiment is
formulated as follows:
\begin{align}
\label{prob:aoed1}
\begin{split}
\min_{\text{Experiment}} & \text{Experimental effort}\\
 s.t.  \; \ \ \ \ &\hat{G}\in \mathcal{E}_{adm}
\end{split}
\end{align}
where $\hat{G}$ is the estimated model. For the experimental
effort, different measures commonly used are input or output
power, and experimental length. For $\hat{G}$, standard
maximum likelihood (ML) and Bayesian estimation methods, e.g., minimum mean square error (MMSE), are usually employed.

Optimizing the experiment
and optimally choosing the system estimator are two problems that should ultimately be tackled in a joint context. Nevertheless, both in the framework of classical and
application-oriented experiment designs, a \emph{separation} strategy is applied: initially, we select and fix the system estimator to a choice that is known
to possess some optimality aspects, e.g., the ML or MMSE estimators, and then we are optimizing the experiment. For finite-length experiments the MMSE estimator is often considered to be superior to the ML estimator. A related meaningful question in the application-oriented framework is: Are there end-performance metrics for which
the ML estimator outperforms the MMSE when the experiment is finite-length?

In this paper, we affirmatively answer the last question based on a simple linear Gaussian regression model that is used here as the simplest possible example to provide the necessary answer. The reason for choosing this example is two-fold: except for the simplicity that it allows, it neutralizes the choice of the optimal experiment. Via this example, we re-examine the validity of the common belief
that the MMSE estimator is superior to the ML estimator, when finite length experiments are used to identify the unknown system. To this end,
appropriate mean square error (MSE)-like end-performance metrics are used that are meaningful is certain applications such as in communication and control systems. Finally, we numerically demonstrate the validity of the claims verifying the purchased analysis.

This paper is organized as follows: Section~\ref{sec:ProbSt}
defines the problem of designing the system estimator with
respect to the end performance metric. Section~\ref{sec:prelim}
presents some results and comments that will be useful in the rest
of the paper, while it introduces approximations of the
performance metrics that the rest of the analysis will be based
on. The optimality of the ML and MMSE system estimators with
respect to the minimization of the aforementioned MSE-like end-performance metrics is examined in Section~\ref{sec:eMSE}. Section~\ref{sec:sims} illustrates the
validity of the derived results. Finally, Section~\ref{sec:concls} concludes
the paper.

\emph{Notations}:  Vectors are denoted by bold letters. Superscripts $^{T}$ and $^{H}$ stand for transposition and Hermitian transposition, respectively. $|\cdot|$ is
the complex modulus. For a vector $\g{a}$, $a(m)$ denotes its \emph{m-th} entry. The expectation operator is denoted by $E(\cdot)$. Finally, $\mathcal{CN}(\mu,\sigma^2)$ denotes the complex Gaussian distribution with mean $\mu$ and variance $\sigma^2$.

\section{Problem Statement}
\label{sec:ProbSt}

Consider the scalar linear Gaussian model
\begin{equation}
y(n)=\theta u(n)+e(n), \label{eq:sm}
\end{equation}
where $y(n)$ is the observed signal at time
instant $n$, $\theta$ is the unknown system parameter assumed to be complex-valued, $u(n)$ is the input at the same time
instant and
$e(n)$ is
complex, circularly symmetric, Gaussian noise with zero mean and
variance $\sigma_e^2$. We further assume that $E[u(n)]=0$ and
$E[|u(n)|^2]=\sigma_u^2$. In addition, $w(n)$ and $u(n)$ are independent random
sequences, while $e(n)$ is a white random sequence.

Assume that the experimental length is limited to $N$ time slots and that the
maximum allowed input energy for experimental purposes is $\mathcal{E}$. We can collect the received samples corresponding to the experiment in
one vector:
\begin{equation}
\gss{y}{\rm exp}{}=\theta\gss{u}{\rm exp}{}+\gss{e}{\rm exp}{},
\label{eq:rxTr}
\end{equation}
where $\gss{y}{\rm exp}{}=\left[y(l-N+1), y(l-N+2),\cdots, y(l)
\right]^{T}$ is the vector of $N$ received samples corresponding
to the experiment, $\gss{u}{\rm exp}{}=\left[ u(l-N+1), u(l-N+2), \cdots,
u(l) \right]^{T}$ is the vector of $N$ input symbols and
$\gss{e}{\rm exp}{}=\left[e(l-N+1), e(l-\right.$ $\left. N+2), \cdots, e(l) \right]^{T}$
is the vector of $N$ noise samples. Considering the class of
linear parameter estimators, the system is estimated as follows:
\begin{equation}
\hat{\theta}=\gss{f}{}{H}\gss{y}{\rm exp}{}=\theta\gss{f}{}{H}\gss{u}{\rm
exp}{}+\gss{f}{}{H}\gss{e}{\rm exp}{}, \label{eq:chEst}
\end{equation}
where $\g{f}$ is a $N\times 1$ estimating filter.

A possible performance metric is the MSE of a
\emph{linear} input estimator. The input estimator uses the
system knowledge and delivers an estimate of the input variable. We call \emph{clairvoyant}
the input estimator that has perfect system knowledge. Denoting the corresponding
estimating filter by $\tilde{c}(\theta)$, we can find its mathematical
expression as follows:
\begin{equation}
\tilde{c}(\theta)=\arg\min_{c(\theta)}
E\left[\left|c(\theta)y(n)-u(n)\right|^2\right], \label{eq:MMSEchoice}
\end{equation}
where the expectation is taken over the statistics of $u(n)$ and
$e(n)$. If we set the derivative of the last expression with
respect to $c(\theta)$ to zero and we solve for $c(\theta)$, then the
optimal clairvoyant input estimating filter is given by the expression
\begin{equation}
\tilde{c}(\theta)=\frac{\sigma_u^2 \theta^{*}}{|\theta|^2\sigma_u^2+\sigma_e^2}.
\label{eq:MMSEeq}
\end{equation}
We will call this the MMSE clairvoyant input estimator\footnote{The multiplication by $y(n)$ is considered implicit.}. We observe that
as the signal-to-noise ratio (SNR) increases, i.e., $\sigma_{e}^2\rightarrow 0$,
$\tilde{c}(\theta)\rightarrow 1/\theta$. We call $\check{c}(\theta)=1/\theta$ the \emph{Zero Forcing}
(ZF) clairvoyant input estimator. Due to this last convergence and for simplicity purposes, we focus only on the
ZF input estimator in the sequel.

We can now introduce an end-performance metric of interest, which will be used in the following analysis.
Given an input estimator, we define the excess of the
input estimate based on an input estimator that only knows a system
estimate over the input estimator with perfect system knowledge, thus leading to
\begin{equation}
{\rm
MSE}_{ex}=E\left[\left|c(\hat{\theta})y(n)-c(\theta)y(n)\right|^2\right].\label{eq:MSEexcess}
\end{equation}
In the sequel, this metric will be called \emph{excess} MSE.

 Our goal will be to determine the optimal parameter estimators
for fixed experiments of finite length so that ${\rm
MSE}_{ex}$ based
on the ZF input estimator is minimized. To this end, the following section
presents some useful ideas.

\section{Preliminary Results}
\label{sec:prelim}

Consider the ML estimator. For the linear Gaussian regression, this estimator coincides with the minimum variance unbiased (MVU) estimator. We therefore replace our references to the ML estimator by references to the MVU estimator from now on. Since the MVU is an unbiased estimator,
it satisfies $\gss{f}{}{H}\gss{u}{\rm exp}{}=1$. This condition
implies that $E[\hat{\theta}]=\theta$. For our problem assumptions, the MVU
estimator can be found by solving the following optimization
problem:
\begin{eqnarray}
&& \min_{\sbm{f}} \sigma_e^2\|\g{f}\|^2\nonumber\\
&& {\rm s.t.}\ \ \gss{f}{}{H}\gss{u}{\rm exp}{}=1.
\end{eqnarray}
Forming the Lagrangian for this problem and zeroing its gradient
with respect to $\gss{f}{}{}$, we get:
\begin{equation}
\gss{f}{\rm MVU}{}=\frac{\gss{u}{\rm exp}{}}{\|\gss{u}{\rm
exp}{}\|^2}.\label{eq:MVUE}
\end{equation}

If we assume that $\theta$ is a random variable and that its prior distribution is known, then
instead of the MVU one could use the MMSE parameter estimator. With
our assumptions and the extra assumption that $E[\theta]=0$, one can
obtain~\cite{k93}
\begin{equation}
\gss{f}{\rm MMSE}{}=\frac{E[|\theta|^2]\gss{u}{\rm
exp}{}}{E[|\theta|^2]\|\gss{u}{\rm exp}{}\|^2+\sigma_e^2}.\label{eq:MMSE}
\end{equation}


Assuming that $\theta$ is a deterministic but unknown variable, the ${\rm MSE}_{ex}$ of the ZF input estimator can be easily obtained:
\begin{equation}
{\rm MSE}_{ex}^{d}\left({\rm
ZF}\right)=E\left[\left|\frac{\hat{\theta}-\theta}{\hat{\theta}}\right|^2\right]\left(\sigma_u^2+\frac{\sigma_e^2}{|\theta|^2}\right)\label{eq:eMSEZF}
\end{equation}
(c.f. (\ref{eq:MSEexcess})). Here, the superscript ``d'' stands for ``deterministic''. If $\theta$ is assumed to be a random variable, then the corresponding end-performance metric ${\rm MSE}_{ex}^{r}$ is
obtained by averaging the last expression over $\theta$.

Depending on the probability distributions of $|\hat{\theta}|$ and $|\theta|$,
the above MSE expressions may fail to exist. The
MSEs will be finite if the probability distribution function (pdf)
of $|\hat{\theta}|$ is of order $O(|\hat{\theta}|^2)$ as $\hat{\theta}\rightarrow
0$. A similar condition should hold for the pdf of $|\theta|$ in the
case of ${\rm MSE}_{ex}^{r}$. In the opposite case, we end up with an \emph{infinite moment} problem. In order to obtain well-behaved parameter estimators that will
be used in conjunction with the actual performance metric, some
sort of regularization is needed. Some ideas for appropriate
regularization techniques to use may be obtained by modifying
robust estimators (against heavy-tailed distributions), e.g., by
trimming a standard estimator, if it gives a value very close to
zero \cite{hu05}. An example of such a trimmed estimator is given as follows:
\begin{eqnarray}
\hat{\theta}= \left\{%
\begin{array}{c}
  \gss{f}{}{H}\gss{y}{\rm exp}{},\ \ {\rm if}\ \ |\gss{f}{}{H}\gss{y}{\rm exp}{}|>\lambda\\
  \lambda
\gss{f}{}{H}\gss{y}{\rm exp}{}/|\gss{f}{}{H}\gss{y}{\rm exp}{}|,\ \ {\rm o.w.} \\
\end{array}
\right.\label{eq:WellBehEst}
\end{eqnarray}
where $\g{f}$ can be any estimator and $\lambda$ a regularization
parameter\footnote{This parameter can be tuned via
cross-validation or any other technique, although in the
simulation section we empirically select it for simplicity
purposes.}.

\emph{Remark:} Clearly, the reader may observe that the definition
of the trimmed $\hat{\theta}$ preserves the continuity at
$|\gss{f}{}{H}\gss{y}{\rm exp}{}|=\lambda$. Additionally, the event
$\{\gss{f}{}{H}\gss{y}{\rm exp}{}=0\}$ has zero probability since
the distribution of $\gss{f}{}{H}\gss{y}{\rm exp}{}$ is continuous.
Therefore, in this case $\hat{\theta}$ can be arbitrarily defined,
e.g., $\hat{\theta}=\lambda$.

Assume a fixed $\lambda$.
Then,
for a sufficiently small $\lambda$ and a sufficiently high SNR during training,
minimizing ${\rm MSE}_{ex}^{d}({\rm ZF})$ is approximately equivalent to
minimizing the approximation
\begin{equation}
\left[{\rm MSE}_{ex}^{d}\left({\rm
ZF}\right)\right]_0=\frac{E\left[\left|\hat{\theta}-\theta\right|^2\right]}{E\left[\left|\hat{\theta}\right|^2\right]}\left(\sigma_u^2+\frac{\sigma_e^2}{|\theta|^2}\right),\label{eq:MSEddcZF0}
\end{equation}
as we show in the appendix.
Using some minor additional
technicalities, we can work with
\begin{align}
&\left[{\rm MSE}_{ex}^{r}\left({\rm
ZF}\right)\right]_0=\nonumber\\& \frac{\sigma_u^2E_{\theta}\left[|\theta|^2E\left[\left|\hat{\theta}-\theta\right|^2\right]\right]+\sigma_e^2E_{\theta}\left[E\left[\left|\hat{\theta}-\theta\right|^2\right]\right]}{E_{\theta}\left[|\theta|^2E\left[|\hat{\theta}|^2\right]\right]},\label{eq:MSEdrcZF0}
\end{align}
instead of ${\rm MSE}_{ex}^{r}\left({\rm ZF}\right)$. We call the last approximations
\emph{zeroth order} input estimate excess MSEs. The following analysis and results will be based on
the zeroth order  metrics and they will reveal the dependency
of the system estimator's selection on the considered (any) end-
performance metric.

\emph{Remarks:}
\begin{enumerate}
\item A useful, alternative way to consider the zeroth order MSEs is to view
them as affine versions of normalized parameter MSEs, where the
actual true parameter is $\hat{\theta}$ and the estimator is $\theta$.
\item In the definition of (\ref{eq:MSEddcZF0}), one can observe that after approximating the mean value of the ratio by the ratio of the mean values the infinite moment problem is eliminated. In the following, all zeroth order metrics will be defined based on the \emph{non-trimmed} $\hat{\theta}$ to ease the derivations. This treatment is approximately valid when $\lambda$ is sufficiently small.
\end{enumerate}

\section{Minimizing the Zeroth Order Excess MSE}
\label{sec:eMSE}

In this section, we investigate the selection of the system estimator
for the zeroth order excess MSE in the case of the ZF input estimator.

\subsection{ZF Input Estimator with a Deterministic System}
\label{subsec:eMSEdcZF}

The expectation operators in Eq. (\ref{eq:MSEddcZF0}) are with
respect to $\gss{e}{\rm exp}{},u(n)$ and $e(n)$.
In this case, we have:
\begin{equation}
\left[{\rm MSE}_{ex}^{d}\left({\rm ZF}\right)\right]_0=
\frac{|\theta|^2\left|\sbm{f}^{H}\sbms{u}{\rm
exp}-1\right|^2+\sigma_e^2\left\|\sbm{f}\right\|^2}{|\theta|^2\left|\sbm{f}^{H}\sbms{u}{\rm
exp}\right|^2+\sigma_e^2\left\|\sbm{f}\right\|^2}\left(\sigma_u^2+\frac{\sigma_e^2}{|\theta|^2}\right)
\end{equation}
The numerator of the gradient of the above expression with respect
to\footnote{discarding the positive scalars and considering again
the corresponding (hermitian) transpositions.} $\gss{f}{}{}$ is
given by the following expression:
\begin{eqnarray}
&&\left[|\theta|^2|\varphi|^2+\sigma_e^2\|\g{f}\|^2\right]\left[|\theta|^2\left(\varphi-1\right)^{*}\gss{u}{\rm
exp}{}+\sigma_e^2\g{f}\right]\nonumber\\
&&-\left[|\theta|^2\varphi^{*}\gss{u}{\rm
exp}{}+\sigma_e^2\g{f}\right]\left[|\theta|^2\left|\varphi-1\right|^2+\sigma_e^2\|\g{f}\|^2\right],\nonumber\\
\label{eq:eZFnom}
\end{eqnarray}
where $\varphi=\gss{f}{}{H}\gss{u}{\rm exp}{}$.
Setting $\g{f}=\gss{f}{\rm MVU}{}$, one can easily check that the
above expression becomes zero. Therefore:

\begin{prop}
The MVU \emph{is} an optimal system estimator for the task of
minimizing $\left[{\rm MSE}_{ex}^{d}\left({\rm
ZF}\right)\right]_0$, when the system parameter is considered a deterministic
but otherwise unknown quantity.
\end{prop}
 \emph{Remark:} Note that even if $\left[{\rm MSE}_{ex}^{
d}\left({\rm ZF}\right)\right]_0$ depends on the unknown system parameter
$\theta$, the optimal system estimator does not in this case.

\subsection{ZF Input Estimator with a Random System}
\label{subsec:eMSErcZF}

In this case, the prior statistics of $\theta$ are known. The
zeroth order excess MSE is given by:
\begin{eqnarray}
\left[{\rm MSE}_{ex}^{r}(ZF)\right]_0&=&\frac{\left|\varphi-1\right|^2(E[|\theta|^4]\sigma_u^2+E[|\theta|^2]\sigma_e^2)}{E[|\theta|^4]|\varphi|^2+\sigma_e^2\left\|\sbm{f}\right\|^2
E[|\theta|^2]}\nonumber\\
&&+\frac{\sigma_e^2\left\|\sbm{f}\right\|^2(E[|\theta|^2]\sigma_u^2+\sigma_e^2)}{E[|\theta|^4]|\varphi|^2+\sigma_e^2\left\|\sbm{f}\right\|^2
E[|\theta|^2]}
\end{eqnarray}
Differentiating this expression w.r.t. $\gss{f}{}{}$ and setting
$\g{f}=\gss{f}{\rm MVU}{}$ we zero the gradient. Therefore:
\begin{prop}
The MVU \emph{is} an optimal system estimator for the task of
minimizing $\left[{\rm MSE}_{ex}^{r}(ZF)\right]_0$, when the
system parameter is considered random.
\end{prop}

Via tedious calculations, we can show that the MMSE channel
estimator does not zero the gradient.

\emph{Remark:} This result is \emph{counterintuitive}: it says that
    when one has knowledge of the system statistics but uses a ZF
    input estimator, one should ignore these statistics in choosing a
    system estimator for minimizing the zeroth order excess MSE. This is the
    \emph{major} result in this paper: The belief that combining the MMSE system estimator
    with any performance metric is better than using the MVU/ML system estimator when finite length
    experiments are used to identify the system, is \emph{not} valid.

\subsection{Discussion on the Optimal Training}

Since the system estimator is selected in order to optimize the
final performance metric, one may
consider the problem of selecting optimally the input vector
$\gss{u}{\rm exp}{}$ under a maximum energy constraint $\|\gss{u}{\rm exp}{}\|^2\leq \mathcal{E}$ to serve the same purpose. To optimize the
input vector, one should first fix the system estimator. This
is a ``complementary'' problem with respect to the approach that we have
followed so far. Suppose that we use either the MVU or the MMSE
system estimators. One can observe that for $N=1$ the problem of
selecting optimally the input vector is meaningless.
In the case that $N>1$, fixing for example $\g{f}=\gss{f}{\rm
MVU}{}$ one can observe that again the problem of selecting
optimally the input vector is meaningless. Consider for example
the case of $\left[{\rm MSE}_{ex}^{r}\left({\rm
ZF}\right)\right]_0$. We then have:
\[
\left[{\rm MSE}_{ex}^{r}\left({\rm
ZF}\right)\right]_0=\frac{\sigma_e^2\left(E[|\theta|^2]\sigma_u^2+\sigma_e^2\right)}{E[|\theta|^4]\|\gss{u}{\rm exp}{}\|^2+\sigma_e^2E[|\theta|^2]},
\]
which only depends on $\|\gss{u}{\rm exp}{}\|^2$.
Furthermore, $\left[{\rm MSE}_{ex}^{r}\left({\rm ZF}\right)\right]_0$ is
minimized when $\|\gss{u}{\rm exp}{}\|^2=\mathcal{E}$, which is
intuitively appealing. Therefore, any $\gss{u}{\rm exp}{}$ with energy
equal to $\mathcal{E}$ is an equally good input vector for the
MVU estimator. Thus, for the same $\gss{u}{\rm exp}{}$, the MVU estimator
is better than the MMSE.

\section{Simulations}
\label{sec:sims}

\begin{figure}
\begin{center}
  \includegraphics[scale=0.5]{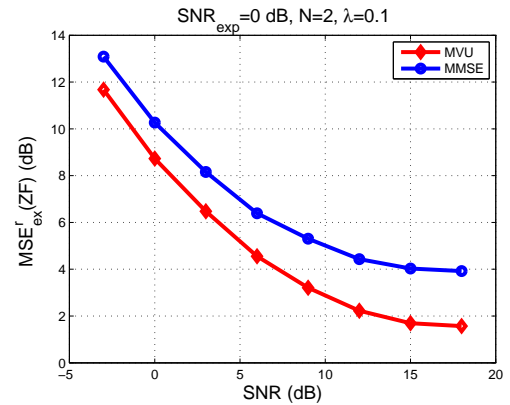}
\end{center}
\caption{${\rm MSE}_{ex}^{r}(\rm ZF)$ with SNR during the experiment
equal to $0$~dB, $N=2$ and $\lambda=0.1$.}
  \label{fig:ZFexMSE}
\end{figure}

In this section we present numerical results to verify our
analysis. In all figures, $\theta\sim\mathcal{CN}(0,1)$. The SNR during the experiment highlights how good
the system estimate is. The parameter $\lambda$ has
been empirically selected to be $0.1$ in Fig.~\ref{fig:ZFexMSE}. The two figures that we present in this section
aim at two goals: first, to highlight that indeed the MVU/ML estimator can be better than the MMSE in finite length
system identification depending on the end-performance metric of interest. And second, to verify that the zeroth order approximations used
in this paper for analysis purposes are good approximations to the true end-performance metrics for extracting the necessary conclusions.

\begin{figure}
\begin{center}
  \includegraphics[scale=0.5]{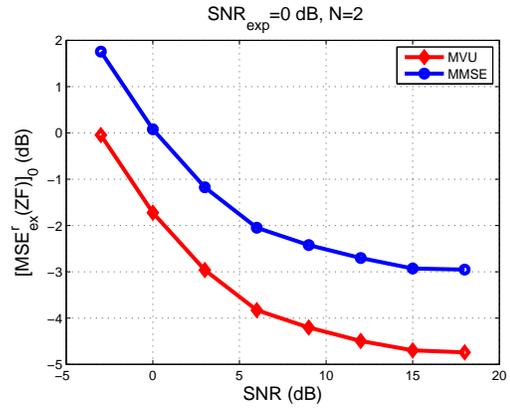}
\end{center}
\caption{$\left[{\rm MSE}_{ex}^{r}(\rm ZF)\right]_0$ with SNR
during the experiment equal to $0$~dB and $N=2$.}
  \label{fig:ZFexMSE_zeroth}
\end{figure}

Fig.~\ref{fig:ZFexMSE_zeroth} presents the corresponding results
for $\left[{\rm MSE}_{ex}^{r}(\rm ZF)\right]_0$. The SNR during the experiment has been set to $0$ dB, which can be a low operational value in real world appplications, but useful, e.g., in situations where energy efficiency is crucial such as in wireless sensor networks. The experimental length has been set to $2$ simply to eliminate the asymptotic efficiency of the ML estimator. The MVU is the
best estimator as proven. This is an example contradicting
what one would expect and verifying the motivation of this paper.

Finally, Fig.~\ref{fig:ZFexMSE} verifies that the zeroth order metrics
used in this paper are good approximations in terms of indicating the structure of uniformly better
estimators than the MMSE. The SNR during the experiment and the experimental length are as before. We observe that except for a translation in the vertical direction, the zeroth order approximations are able to indicate the relative position of the estimating curves leading to accurate conclusions about the
comparison between them.

\section{Conclusions}
\label{sec:concls}

In this paper, end-performance metric system estimator selection
has been investigated. We have shown that the application-oriented
selection is the right way to choose estimators in practice. We
have verified this observation based on an explanatory end-performance metric of interest, namely, the excess input estimate MSE. The extracted conclusion
is that the ML/MVU estimators can be better than the MMSE estimator for particular end-performance metrics of interest. This invalidates the common belief that the MMSE estimator is always better than the ML/MVU estimators for \emph{any} end-performance metric, if
finite length experiments are used for system identification purposes.

\appendix

This section proposes a simplification of the $\left[{\rm MSE}_{ex}^{d}\left({\rm
ZF}\right)\right]$ metric for the estimator given in
\pref{eq:WellBehEst} with a fixed $\lambda$. Due to the
Gaussianity of $\gss{u}{\rm exp}{}$, $\left[{\rm MSE}_{ex}^{d}\left({\rm
ZF}\right)\right]=\infty$ for any $\g{f}\neq \g{0}$ (infinite moment problem). Using
\pref{eq:WellBehEst}, the corresponding metric becomes:
\begin{eqnarray}
&&\left[{\rm MSE}_{ex}^{d}\left({\rm
ZF}\right)\right]_{\rm reg}={\rm
Pr}\left\{|\gss{f}{}{H}\gss{y}{\rm
exp}{}|>\lambda\right\}\cdot\nonumber\\
&&E\left[\left(\sigma_u^2+\frac{\sigma_e^2}{|\theta|^2}\right)\left|1-\frac{\theta}{\gss{f}{}{H}\gss{y}{\rm
exp}{}}\right|^2; |\gss{f}{}{H}\gss{y}{\rm
exp}{}|>\lambda\right]\nonumber\\&& + {\rm
Pr}\left\{|\gss{f}{}{H}\gss{y}{\rm exp}{}|\leq \lambda\right\}
\cdot\nonumber\\
&& E\left[\left(\frac{\sigma_u^2}{\lambda^2}+\frac{\sigma_e^2}{\lambda^2|\theta|^2}\right)\left|\lambda
\frac{\gss{f}{}{H}\gss{y}{\rm exp}{}}{|\gss{f}{}{H}\gss{y}{\rm
exp}{}|}-\theta\right|^2;
|\gss{f}{}{H}\gss{y}{\rm exp}{}|\leq\lambda\right],\nonumber\\
\end{eqnarray}
where $;$ denotes conditioning and ``reg'' signifies the use of the regularized system estimator in \pref{eq:WellBehEst}.
Moreover,
${\rm Pr}\left\{|\gss{f}{}{H}\gss{y}{\rm exp}{}|\leq
\lambda\right\}=O(\lambda^2)$, since by the mean value theorem
this probability is equal to the area of the region
$\{|\gss{f}{}{H}\gss{y}{\rm exp}{}|\leq \lambda\}$, which is of
order $O(\lambda^2)$, multiplied by some value of the probability
density function of $|\gss{f}{}{H}\gss{y}{\rm exp}{}|$ in that
region, which is of order $O(1)$. In addition,
\begin{eqnarray}
&&E\left[\left(\frac{\sigma_u^2}{\lambda^2}+\frac{\sigma_e^2}{\lambda^2|\theta|^2}\right)\left|\lambda
\frac{\gss{f}{}{H}\gss{y}{\rm exp}{}}{|\gss{f}{}{H}\gss{y}{\rm
exp}{}|}-\theta\right|^2;
|\gss{f}{}{H}\gss{y}{\rm
exp}{}|\leq\lambda\right]=\nonumber\\ &&\left(\sigma_u^2+\frac{\sigma_e^2}{|\theta|^2}\right)
+\left(\frac{\sigma_u^2}{\lambda^2}|\theta|^2+\frac{\sigma_e^2}{\lambda^2}\right)\nonumber\\&&-2\left(\frac{\sigma_u^2}{\lambda}+\frac{\sigma_e^2}{\lambda|\theta|^2}\right)E\left[\Re\left\{
\theta^{*}\frac{\gss{f}{}{H}\gss{y}{\rm exp}{}}{|\gss{f}{}{H}\gss{y}{\rm
exp}{}|}\right\}\right].\nonumber
\end{eqnarray}
Furthermore, if the SNR during training is sufficiently high and the probability mass of $|\gss{f}{}{H}\gss{y}{\rm exp}{}|$ is
concentrated around $|\theta|$, then it
can be shown that
\begin{eqnarray}
&& E\left[\left(\sigma_u^2+\frac{\sigma_e^2}{|\theta|^2}\right)\left|1-\frac{\theta}{\gss{f}{}{H}\gss{y}{\rm
exp}{}}\right|^2; |\gss{f}{}{H}\gss{y}{\rm
exp}{}|>\lambda\right] \nonumber\\ && \approx
\frac{(\sigma_u^2+\sigma_e^2/|\theta|^2)E[|\gss{f}{}{H}\gss{y}{\rm
exp}{}-\theta|^2; |\gss{f}{}{H}\gss{y}{\rm
exp}{}|>\lambda]}{E[|\gss{f}{}{H}\gss{y}{\rm
exp}{}|^2; |\gss{f}{}{H}\gss{y}{\rm
exp}{}|>\lambda]}.\nonumber\\
\end{eqnarray}
The same holds even if $\gss{f}{}{H}\gss{y}{\rm exp}{}$ is a biased estimator of $\theta$ at high training SNR and $|\gss{f}{}{H}\gss{y}{\rm exp}{}|$ tends to concentrate
around a value $\beta$ bounded away from $|\theta|$ (and of course from $0$).

To show the last claim, we set $X=|\gss{f}{}{H}\gss{y}{\rm exp}{}-\theta|^2$ and $Y=|\gss{f}{}{H}\gss{y}{\rm exp}{}|^2$. Since $Y>\lambda^2$, it also holds that $E\left[Y\right]>\lambda^2$. Furthermore, it can be seen that
\begin{align}
\left|E\left[\frac{X}{Y}\right]-\frac{E[X]}{E[Y]}\right|\leq \frac{1}{\lambda^4}E\left[\left|XE[Y]-YE[X]\right|\right].\label{eq:Stochineq}
\end{align}
At high training SNR, $X\rightarrow E[X]$ and $Y\rightarrow E[Y]$ in the mean square sense and therefore it can be easily shown that the right hand side of (\ref{eq:Stochineq}) converges to $0$. To see this, notice that the Cauchy-Schwarz inequality yields
\begin{align}\label{eq:Cauchyineq1}
&\frac{1}{\lambda^4}E\left[\left|XE[Y]-YE[X]\right|\right]\leq \frac{1}{\lambda^4}\left(E\left[\left|XE[Y]-YE[X]\right|^2\right]\right)^{1/2}\nonumber\\
&=\frac{1}{\lambda^4}\left(E^2[Y]E[X^2]+E[Y^2]E^2[X]-2E[XY]E[X]E[Y]\right)^{1/2}.
\end{align}
Since $X\rightarrow E[X]$ and $Y\rightarrow E[Y]$ in the mean square sense, $E[X^2]\rightarrow E^2[X]$, $E[Y^2]\rightarrow E^2[Y]$ and $E[XY]\rightarrow E[X]E[Y]$.
For the last case, notice that
\begin{align*}\label{eq:Cauchyineq2}
&\left|E[XY]-E[X]E[Y]\right|\leq  \sqrt{E\left[\left|X-E[X]\right|^2\right]E\left[\left|Y-E[Y]\right|^2\right]},
\end{align*}
where the last inequality follows again from the Cauchy-Schwarz inequality.
By the mean square convergence of $X$ to $E[X]$ and $Y$ to $E[Y]$ the right hand side of the last inequality tends to $0$.
Therefore, the right hand side of (\ref{eq:Cauchyineq1}) tends to $0$.

Moreover, under the high SNR assumption the conditional
expectations can be approximated by their unconditional ones,
since for a sufficiently small $\lambda$ their difference is due to an event of probability
$O(\lambda^2)$. Therefore,
\begin{eqnarray}
&& E\left[\left(\sigma_u^2+\frac{\sigma_e^2}{|\theta|^2}\right)\left|1-\frac{\theta}{\gss{f}{}{H}\gss{y}{\rm
exp}{}}\right|^2|; |\gss{f}{}{H}\gss{y}{\rm
exp}{}|>\lambda\right]\approx\nonumber\\ &&
\left\{\frac{(\sigma_u^2+\sigma_e^2/|\theta|^2)E[|\gss{f}{}{H}\gss{y}{\rm
exp}{}-\theta|^2]}{E[|\gss{f}{}{H}\gss{y}{\rm
exp}{}|^2]}\right\}+O(\lambda^2).
\end{eqnarray}
Combining all the above results yields
\begin{equation}\label{eq:MSExdcApp}
\left[{\rm MSE}_{ex}^{d}({\rm ZF})\right]_{\rm reg}\approx\left\{\frac{(\sigma_u^2+\sigma_e^2/|\theta|^2)E[|\gss{f}{}{H}\gss{y}{\rm exp}{}-\theta|^2]}{E[|\gss{f}{}{H}\gss{y}{\rm
exp}{}|^2]}\right\}+O(1).
\end{equation}
The $O(1)$ term is not negligible but for sufficiently small $\lambda$ its dependence on $\g{f}$ is insignificant. Hence,
for a sufficiently small $\lambda$ and a sufficiently high SNR during training,
minimizing $\left[{\rm MSE}_{ex}^{d}({\rm ZF})\right]_{\rm reg}$ is equivalent to
minimizing (\ref{eq:eMSEZF}).

\addtolength{\textheight}{-12cm}   

\end{document}